\documentclass{aa}
\usepackage[varg]{txfonts}
\usepackage{xcolor}
\usepackage{siunitx}
\usepackage{ulem}
\usepackage{upgreek}

\bibpunct{(}{)}{;}{a}{}{,} 

\providecommand{\kms}{\ensuremath{\rm \,km\,s^{-1}}\xspace}
\providecommand{\Msun}{\ensuremath{\,{\cal M}_{\odot}}\xspace}
\providecommand{\Ftwo}{\ensuremath{\mathrm{F2}}\xspace}

\begin{document}

\title{Binaries masses and luminosities with Gaia DR3}
\author{S. Chevalier
        \inst{1,2,3}
        \and
        C. Babusiaux
        \inst{1}
        \and
        T. Merle
        \inst{4,5}
        \and 
        F. Arenou
        \inst{6}
        }

\institute{Univ. Grenoble Alpes, CNRS, IPAG, 38000 Grenoble, France
            \and Kungliga Tekniska Högskolan, 114 28 Stockholm, Sverige
            \and Ecole Centrale de Lyon, 69 130 Ecully, France
            \and Institut d’Astronomie et d’Astrophysique, Université Libre de Bruxelles (ULB), Belgium
            \and Royal Observatory of Belgium, Avenue Circulaire 3, 1180 Brussels, Belgium
            \and GEPI, Observatoire de Paris, Universit{\'e} PSL, CNRS, 92190 Meudon, France
}

\date{Received / Accepted }

\abstract
   {The recent Gaia third data release (DR3) has brought some new exciting data about stellar binaries. It provides new opportunities to fully characterize more stellar systems and contribute to enforce our global knowledge of stars behaviour.} 
   {By combining the new Gaia non-single stars catalog with double-lined spectroscopic binaries (SB2), one can determine the individual masses and luminosities of the components. To fit an empirical mass-luminosity relation in the Gaia $G$ band, lower mass stars need to be added. Those can be derived using Gaia resolved wide binaries combined with literature data. 
   }
   {
   Using the BINARYS tool, we combine the astrometric non-single star solutions in the Gaia DR3 with SB2 data from two other catalogs : the 9th Catalogue of Spectroscopic Binary orbits (SB9) and APOGEE. We also look for low mass stars resolved in Gaia with direct imaging and Hipparcos data or literature mass fraction.  }
   {The combination of Gaia astrometric non-single star solutions with double-lined spectroscopic data enabled to characterize 43 binary systems with SB9 and 13 with APOGEE. We further derive the masses of 6 low mass binaries resolved with Gaia. We then derive an empirical mass-luminosity relation in the Gaia $G$ band down to 0.12\Msun.}
  {}


\titlerunning{Gaia DR3 Mass Catalog}
\authorrunning{S. Chevalier}
\maketitle

\section{Introduction} \label{introduction}

The $3^{\rm rd}$ data release (DR3) from the Gaia mission \citep{GDR3summary} provides for the first time non-single star solutions for hundreds of thousands sources \citep{DR3NSS}. This brings a new exciting dataset to fully characterize new binary systems in particular their dynamical masses and luminosities.

The estimation of stellar masses is a fundamental process to improve the understanding of their behaviour (luminosity, evolution, etc.). This can be mainly achieved by characterizing binary systems, which is the aim of this paper. The stars that can be fully characterized like the ones studied in this paper are not that many, but they are crucial since they enable to calibrate fundamental physical relations. These ones will then make possible to estimate the parameters of single stars or less reachable objects. The main example is the mass-luminosity relation.
It is set on fully characterized star systems, for which masses and luminosities of both components are known. Knowing the dynamical masses also enables to constrain other characteristics of the stars, such as their age through isochrone fitting.  

One of the main purpose of this paper is to use these new Gaia DR3 data to 
provide new dynamical masses and a first mass luminosity relation in the G band. Empirical mass luminosity relations are mostly provided in the near-infrared due to its lower dependency on the metallicity than the visible \citep[e.g.][]{ML_delfosse, ML_benedict, MLrelation}. In the visible empirical mass luminosity relations are provided in the V band \citep[e.g.][]{ML_delfosse, ML_benedict}. 

Masses of double-lined spectroscopic binaries (SB2) have been obtained so far mainly through eclipsing binaries and a smaller sample of visually resolved binaries, the latter having the advantage of providing also a measure of the parallax of the system \citep[e.g.][]{Pourbaix00}.  Masses can also be estimated together with the luminosity of the stars through the astrometric motion of the photocentre. However this motion was in general too small to be detected by Hipparcos \citep{HIP} \citep[see e.g.][]{Jancart-SB9orbits}, except in a few cases \citep{2000IAUS..200P.135A}. 
An observing program of SB2 has been initiated since 2010 to allow the determination of masses at the 1\% level using future Gaia astrometric orbits \citep{2020MNRAS.496.1355H}.
The new Gaia DR3 astrometric orbits allow the determination of new masses of SB2 systems, as already done between the Gaia SB2 and astrometric orbital solutions in \cite{DR3NSS}. 
However the astrometric motion impacted the epoch radial velocity measures of Gaia \citep{DR3Validation}, leading to bad goodness of fit of the solutions. Those will therefore not be considered here.
Here we combine Gaia astrometric data with double-lined spectroscopic data from APOGEE \citep[Apache Point Observatory Galactic Evolution Experiment,][]{APOGEESB2} and the 9th Catalogue of Spectroscopic Binary orbits \citep[SB9,][]{SB9} to derive the dynamical mass of each component as well as their flux in the $G$ band.

Section \ref{data-presentation} presents the data used: the double-lined spectroscopic data (Sect.~ \ref{data-presentation-SB2}) and the astrometric solutions from Gaia (Sect.~\ref{data-presentation-Gaia}). 
Then the method used to determine the binary masses is explained in Sect.~\ref{data-process}.
The results obtained are discussed in Sect.~\ref{results}. 
Section \ref{ML-relation} is dedicated to the mass-luminosity relation, presenting first 6 low-mass stars resolved by Gaia with direct imaging data (Sect.~\ref{sec:resolvedstars}) and then the fit of mass-luminosity relation (Sect.~\ref{sec:mlfit}). 

\section{Data} \label{data-presentation}

We determine binary system features using astrometric data combined with double-lined spectroscopy.


\subsection{The double-lined spectroscopic data} \label{data-presentation-SB2}

Spectroscopic data has been obtained by measuring the Doppler effect of the star system. The stellar motion induces a periodic translation of their spectrum depending on their motion in the line-of-sight from the Earth. For double-lined spectroscopy, the motion of the two sources of the binary are well identified and the ratio of the amplitude of the radial velocity motion provides the mass ratio of the star system. The estimation of the mass requires the knowledge of the inclination which cannot be obtained from the spectroscopic orbit. 

We used double-lined spectroscopic data originating from two catalogs: the Apache Point Observatory Galactic Evolution Experiment (APOGEE) and the 9th Catalog of Spectroscopic Orbits (SB9).  \\

SB9 \citep[][]{SB9} is a huge compilation of spectroscopic orbits from the literature over the past decades. It counts about 5000 orbits together with the input radial velocities used to derive the orbit. From those, 55 have epoch radial velocities available and a Gaia astrometric orbit counterpart.
The compatibility between the orbital solutions provided by Gaia astrometry and SB9 spectroscopy has been checked first to remove triple systems. We required a consistency at $10\sigma$ for the periods $P$ :
\begin{equation} \label{eq-period-compat1}
    \left| P_{Gaia} - P_{SB9} \right| < 10 \sqrt{\sigma_{P_{Gaia}}^2 + \sigma_{P_{SB9}}^2}.
\end{equation}
We note that a consistency at $5\sigma$ would have removed the well behaved solution Gaia DR3 1528045017687961856 (HIP~62935).
The compatibility of the eccentricity within $10~\sigma$ was also checked (like in Eq.~\ref{eq-period-compat1}) but does not remove any star system, leading to 43 SB9 binaries selected. 

APOGEE \citep[][]{APOGEE} is a survey conducted with two high resolution spectrographs, covering the spectral band between $1.51$ and $1.7$ $\upmu$m. Data used in this paper originates from \cite{APOGEESB2}, in which 7273 double-lined spectroscopic systems have been detected. 
183 star systems have been found to have both double-lined SB2 spectroscopy data in APOGEE and an astrometric orbital solution in Gaia non-single star (NSS) catalog, but only 126 that had more than one APOGEE epoch have been kept. 

Among them, only one star system orbit could be solved through spectroscopy only by \cite{APOGEESB2}, Gaia DR3 $702393458327135360$ (HD~80234). The direct combination of those spectroscopic parameters with Gaia astrometric ones has already been achieved by \cite{DR3NSS} to derive the masses of this system. 
For the other stars with a Gaia NSS solution counterpart, the constrains from the astrometric orbit can be used to extract the mass ratio from the raw radial velocity curves. \\

SB9 radial velocity curves have a much larger observation time range than APOGEE. There are enough data from various epochs to perform an independent fit of the orbit with spectroscopy only. The orbital parameters 
are directly given with their associated errors in the catalog. This is not the case for our APOGEE sample, except for Gaia DR3 $702393458327135360$ (HD~80234).

\subsection{The Gaia DR3 astrometric orbits} \label{data-presentation-Gaia}

Astrometric data is obtained by observing the corkscrew-like motion of the photocentre i.e. the apparent light source of the binary system. 
It provides the needed inclination but also the parallax and, combined with an SB2 solution, the flux ratio.

Here the astrometric data is given by the orbital solutions from the Gaia $3^{\rm rd}$ data release non-single star solutions catalog. 
This new catalog enabled to increase significantly the number and the precision of binary system solutions \citep{DR3NSS}. The catalog provides several types of solutions depending on the collected data and the detection method or instrument used, that is Eclipsing, Spectroscopic and Astrometric solutions and potential combination of those. In this paper, we only use the astrometric solutions. Among our final sample, 21 have a combined AstroSpectroSB1 solution for which only the astrometric part is taken into account. 

In Gaia DR3, the orbit is not described by the Campbell elements (semi-major axis, inclination, node angle, periastron angle) but by the Thiele-Innes coefficients \citep{NSSastro}.

\section{Data processing}  \label{data-process}

While for SB9 the orbital parameters are known and the computation of the masses can be derived directly (Annex \ref{annex-direct-calculation-sb9}), we go back here to the raw spectroscopic data to improve how the correlations between the parameters are considered.

The combination of spectroscopy and astrometry is achieved using BINARYS \citep[orBIt determiNAtion with Absolute and Relative
astrometRY and Spectroscopy,][]{BINARYS}. BINARYS can combine Hipparcos and/or Gaia absolute astrometric data with relative astrometry and/or radial velocity data. It
has been updated to handle Gaia NSS solutions and its heart which computes the likelihoods is available online\footnote{https://gricad-gitlab.univ-grenoble-alpes.fr/ipag-public/gaia/binarys}. It needs initial values and uses the automatic differentiation code TMB \citep[Template Model Builder, ][]{TMB} to find the maximum likelihood. It gives in output the estimated orbital parameters with the associated covariance matrix together with a convergence flag. Due to the fact that Monte-Carlo techniques cannot be used 
with the astrometric Thiele Innes coefficients of Gaia DR3 \citep[see Section 6.1 of][]{DR3Validation}, the MCMC (Markov Chain Monte Carlo) option of BINARYS cannot be used in this study 
while the TMB automatic differentiation is consistent with the local linear approximation result.

BINARYS provides among all the orbital parameters the primary semi-major axis $a_1$, the mass ratio $q=\mathcal{M}_2/\mathcal{M}_1$ and the period $P$ with their associated covariance matrix. This enables to deduce the primary and secondary masses (see Eq. \ref{primary-mass-jacobian} and \ref{secondary-mass-jacobian}) with the associated errors (Annex \ref{annex-error-jacobian-calculation}). 
\begin{equation} \label{primary-mass-jacobian}
    \mathcal{M}_1 = \frac{a_1^3 ~(1+ q)^2}{P^2 ~q^3}
\end{equation}
\begin{equation} \label{secondary-mass-jacobian}
    \mathcal{M}_2 = \frac{a_1^3 ~(1+ q)^2}{P^2 ~q^2}
\end{equation}
with the period $P$ in years, $a_1$ in au and the masses $\mathcal{M}$ in solar masses \Msun.
It also gives the flux fraction of the secondary $\beta$ in the G spectral band:
\begin{equation} \label{eq-def-beta}
    \beta = \frac{F_2}{F_1+F_2} = \frac{q}{1+q} \left(1 - \frac{a_0}{a_1} \right)
\end{equation}
with $a_0$ the semi-major axis of the photocentre in the same unit as $a_1$ (see Annex \ref{annex-direct-calculation-sb9}).


\subsection{The 9th catalog of spectroscopic orbit - SB9}  \label{data-process-sb9}

BINARYS uses here as input the radial velocity epoch data for each component, the orbital astrometric solution from Gaia NSS and initial parameters, chosen for SB9 to be the result of the direct calculation process (Annex~\ref{annex-direct-calculation-sb9}). 

Inflation of the raw radial velocities uncertainties is quite often needed, either due to an under-estimation of the formal errors, template mismatch or stellar variability effects. We therefore apply a procedure similar to \cite{2020MNRAS.496.1355H} to correct the uncertainties. 

We first apply the variance weight factors $w$ provided in the SB9 database. Those have been provided by some studies combining different observations and gives their relative weights. We therefore start from the weighted uncertainties
 $\sigma = \frac{\sigma_{0}}{\sqrt{w}}$.

Then those uncertainties are adjusted using the goodness of fit estimator \Ftwo \citep{1931PNAS...17..684W}:
\begin{equation} \label{goodness-of-fit}
     \Ftwo = \left(\frac{9 \nu}{2} \right)^{1/2} \left[\left(\frac{\chi^2}{\nu} \right)^{1/3} + \frac{2}{9 \nu} -1 \right]
\end{equation}
with $\nu$ the number of degrees of freedom and $\chi^2$ the weighted sum of the squares of the differences between the predicted and the observed values. The radial velocity uncertainties are scaled to obtain $\Ftwo=0$ \textit{i.e.} $\chi^2 = \chi_0^2$:
\begin{equation} \label{eq-chi20}
    \chi_0^2 = \nu \left( 1 - \frac{2}{9 \nu} \right)^3
\end{equation}
The corrected uncertainties are then
\begin{equation} \label{equation-correction-error}
\sigma_{\rm corr} = \sqrt{\frac{\chi^2}{\chi_0^2}} \times \sigma.
\end{equation}

This correction factor is applied 3 times: the uncertainties are adjusted once independently for each component with a SB1 correction, and then adjusted again together with a SB2 correction over the whole system. 

The process requires the number of degree of freedom to be positive \textit{i.e.} to have more epochs than parameters to fit. This is always the case except for Gaia DR3 $1480959875337657088$ (HIP~69885), which has only 2 epochs for the primary and the secondary. No uncertainty-correction at all is applied for this one. 
In the literature, an orbit fit could have been achieved using additional blended radial velocity epochs which could not be used here. For this star the orbital parameters are mainly driven by the Gaia NSS solution.
Four other star systems do not have enough radial velocity epochs for the secondary to have a SB1 solution necessary to apply the correction process: Gaia DR3 $1441993625629660800$,
 $1517927895803742080$, $4145362250759997952$ and $4354357901908595456$, (respectively HIP~66511, HIP~61436, HD~163336~B and HIP~81170). For them, the two other error correction factors, SB1 primary and SB2, are still applied and the $\chi^2$ of the final solution on the secondary has been checked to be small.

Around 10 star systems have had a significant SB1 correction factor over the primary and/or the secondary, with $\sqrt{\frac{\chi^2}{\chi_0^2}} > 1.3$.

\subsection{The Apache Point Observatory Galactic Evolution Experiment - APOGEE}  \label{data-process-apogee}

Similarly, BINARYS is provided the APOGEE radial velocity epoch data for each component, the orbital astrometric solution from Gaia NSS. But since for APOGEE the spectroscopic orbit is not known, the BINARYS code must be initialized with various initial parameters. We used sampled initial values of $\mathcal{M}_1$ from $0.6$ to $1.4$ solar mass with a $0.2$ step, $\mathcal{M}_2$ from $0.6$ to $1.4$ solar mass with a $0.2$ step (keeping $\mathcal{M}_2 \leq \mathcal{M}_1$), $\beta$ from $0$ to $0.5$ with a $0.1$ step.
Since the direction of motion is set by the spectroscopy, we also try different configurations for the node angle, adding or not a $\pi$ angle to both the node angle $\Omega$ and the argument of periastron $\omega$ to the Gaia astrometric orbit values.
For each system, there is then $15 \times 6 \times 2 = 180$
 initial configurations tested for each star system. \\

The convergence of TMB towards a good solution is not expected for every system: many will be triple systems with the short period binary seen by APOGEE and the longer period one seen by Gaia. 
Each TMB output corresponding to an initial configuration of a given star system must then fit the following  criteria to be kept: it must converge, with a goodness-of-fit estimator $\Ftwo < 5$ 
and the flux fraction of the secondary should be within the interval $[0; ~0.5]$ at $3 ~\sigma$. The star system is kept only if those conditions are met by at least $10 \%$ of the $180$ initial configurations tested.
Then, for each star system, only the solution obtained for more than $80\%$ of the cases where TMB converged is kept. If no such solution exists, the star system is rejected. 

From the 126 star systems studied, 35 remain at this point.
Due to the small number of radial velocity epochs, the solution may have a too low precision on the masses to be interesting despite a good convergence. For the final selection, we keep only stars with $\frac{\sigma_{q}}{q} < 0.5$, $ \frac{\sigma_{\mathcal{M}_1}}{\mathcal{M}_1} < 0.5$ and $\sigma_{\mathcal{M}_1} < 1 \Msun$ leading to 13 systems. 
This selection step is only applied for APOGEE, since the selection over SB9 star systems has been performed through the compatibility of periods.

\section{Results} \label{results}

\begin{figure}[h!]
    \includegraphics[scale=0.5]{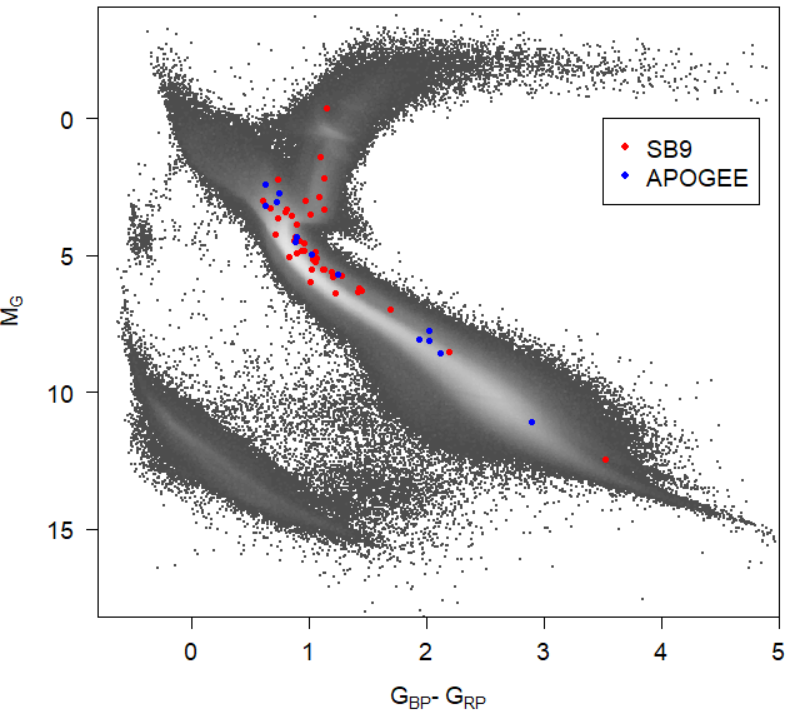}
    \caption{H-R diagram of the characterized binaries. The absolute magnitude of the unresolved binary in the $G$-band $M_G$ is plotted as a function of the colour $G_{BP}-G_{RP}$. The blue dots correspond to the APOGEE binaries, and the red dots correspond to the SB9 binaries. They are overplotted on the low extinction Gaia DR3 HR diagram (in grey).
    }
    \label{hr-diag}
\end{figure}

We have obtained the dynamical masses and flux fraction of the individual stars of 56 binary systems through the combination of Gaia DR3 NSS astrometric solutions with SB2 solutions from SB9 (43 systems) and APOGEE (13 systems). Figure \ref{hr-diag} provides the position of the binaries we have characterized in the HR diagram. 

The results obtained through the orbit fitting process are given in Table \ref{table-orbit-fitting-sb9} for SB9. 

\begin{table*}[!ht]
    \caption{Solutions from the combination of Gaia NSS astrometric solutions with SB9 double-lined spectroscopy. $q$ is the mass ratio, $\mathcal{M}_1$ is the mass of the primary, $\mathcal{M}_2$ is the mass of the secondary (in \Msun) and $\beta$ is the flux fraction of the secondary. $\sigma_q$, $\sigma_{\mathcal{M}_1}$, $\sigma_{\mathcal{M}_2}$ and $\sigma_\beta$ are their associated uncertainties. $\Ftwo$ is the goodness-of-fit estimator. Ref gives the reference of the raw radial velocity data used in the fit, as provided by SB9.}
    \label{table-orbit-fitting-sb9}
    \centering
    \begin{tabular}{lccccccccrr}
        \hline
        Gaia DR3 ID & $q$ & $\sigma_{q}$ & $\mathcal{M}_1$ & $\sigma_{\mathcal{M}_1}$ & $\mathcal{M}_2$ & $\sigma_{\mathcal{M}_2}$ & $\beta$ & $\sigma_{\beta}$ & $\Ftwo$ & Ref \\ \hline \hline 
  48197783694869760&0.6771&0.0124&0.9707&0.0423&0.6573&0.0190&0.0936&0.0030& 1.32&1\\
  69883417170175488&0.9793&0.0224&0.8271&0.0289&0.8100&0.0273&0.3976&0.0084&$-0.03$&2\\
 308256610357824640&0.6970&0.0099&0.4283&0.2000&0.2985&0.1393&0.2770&0.0232&$-0.15$&3\\
 478996438146017280&0.9212&0.0064&0.8112&0.0249&0.7473&0.0228&0.3879&0.0030& 1.25&4\\
 544027809281308544&0.8336&0.0178&0.9465&0.0510&0.7890&0.0317&0.1363&0.0072& 0.01&5\\
 595390807776621824&0.8822&0.0061&0.7877&0.0105&0.6948&0.0066&0.3123&0.0017&$-0.03$&6\\
 660622010858088320&0.8707&0.0367&0.7304&0.0607&0.6360&0.0341&0.2252&0.0081& 1.01&5\\
 827608625636174720&0.9403&0.0129&0.7357&0.0392&0.6917&0.0367&0.4076&0.0045& 0.29&7\\
 882872210352301568&0.5027&0.0026&1.0044&0.0123&0.5049&0.0036&0.0257&0.0038&$-0.07$&8\\
1067685718250692352&0.8884&0.0036&1.0279&0.0125&0.9132&0.0101&0.3035&0.0048& 0.10&9\\
1074883087005896320&0.9388&0.0261&0.6640&0.0393&0.6233&0.0326&0.3977&0.0068& 1.11&10\\
1324699172583973248&0.8839&0.0087&1.3136&0.0476&1.1611&0.0376&0.3167&0.0039& 0.40&11\\
1441993625629660800&0.8344&0.0128&1.1033&0.0455&0.9205&0.0293&0.0706&0.0132&$-0.10$&12\\
1480959875337657088&0.9351&0.0771&0.8679&0.1330&0.8116&0.0925&0.2884&0.0180& 0.77&13\\
1517219363639758976&0.8198&0.0062&1.0650&0.0405&0.8731&0.0313&0.1576&0.0075& 1.55&14\\
1517927895803742080&0.9076&0.0148&0.6354&0.0251&0.5767&0.0183&0.3546&0.0037& 1.75&7\\
1528045017687961856&0.6793&0.0029&1.1179&0.0285&0.7593&0.0185&0.0623&0.0045& 2.30&15\\
1615450866336763904&0.7864&0.0106&1.0138&0.0396&0.7973&0.0261&0.1317&0.0044& 0.14&16\\
1918953867019478144&0.9308&0.0007&1.0465&0.0050&0.9741&0.0047&0.3866&0.0014& 0.63&17\\
2012218158438964224&0.9486&0.0150&2.3368&0.1644&2.2168&0.1509&0.3589&0.0064& 0.92&18\\
2035577729682322176&0.9024&0.0089&0.7027&0.0417&0.6341&0.0374&0.3227&0.0056&$-0.16$&19\\
2067948245320365184&0.7863&0.0006&0.8374&0.0023&0.6584&0.0017&0.1521&0.0014& 0.48&17\\
2129771310248902016&0.9071&0.0183&0.7408&0.0399&0.6720&0.0302&0.3184&0.0048&$-0.16$&10\\
2185171578009765632&0.9629&0.0082&1.2178&0.0299&1.1727&0.0333&0.4436&0.0033& 1.28&17\\
2198442167969655296&0.7508&0.0152&0.9844&0.0416&0.7391&0.0189&0.1963&0.0033& 0.65&10\\
3283823387685219328&0.7419&0.0004&0.9864&0.0112&0.7319&0.0083&0.1478&0.0031& 0.37&20\\
3312631623125272448&0.7062&0.0024&0.9920&0.0304&0.7005&0.0215&0.1249&0.0051& 1.54&21\\
3366718833479009408&0.6407&0.0022&1.0494&0.0102&0.6724&0.0051&0.0155&0.0028& 0.18&8\\
3409686270424363008&0.7076&0.0031&0.7927&0.0139&0.5610&0.0087&0.1086&0.0028& 2.07&17\\
3427930123268526720 \tablefoottext{o} &0.3311&0.0047&0.8066&0.0760&0.2671&0.0248&0.0616&0.0066& 5.23 &3 \\
3536759371865789568&0.8392&0.0212&1.6490&0.0899&1.3838&0.0492&0.2000&0.0061& 0.62&22\\
3549833939509628672&0.9406&0.0195&0.9450&0.0384&0.8889&0.0249&0.3567&0.0051& 0.19&22\\
3931519127529822208&0.7750&0.0260&1.0791&0.0762&0.8363&0.0342&0.1403&0.0045& 1.19&23\\
3935131126305835648&0.7039&0.0022&1.4860&0.0744&1.0460&0.0523&0.1374&0.0052& 0.08&20\\
3954536956780305792&0.9053&0.0094&0.7713&0.7626&0.6982&0.6904&0.3262&0.0600& 0.49&5\\
3964895043508685312&0.8650&0.0084&0.7531&0.1654&0.6514&0.1426&0.2201&0.0196&$-0.09$&17\\
4145362250759997952&0.6960&0.0216&0.9414&0.0611&0.6552&0.0261&0.0757&0.0046& 0.16&24\\
4228891667990334976&0.8629&0.0057&1.0907&0.0588&0.9412&0.0502&0.2872&0.0038& 4.10&25\\
4354357901908595456&0.5832&0.0194&0.7005&0.0509&0.4085&0.0177&0.0606&0.0041& 0.18&26,27\\
4589258562501677312&0.8201&0.0317&0.6679&0.0628&0.5478&0.0395&0.2924&0.0085&$-0.13$&5\\
5762455439477309440&0.8041&0.0327&0.8058&0.0674&0.6480&0.0333&0.1768&0.0076& 0.20&5\\
6244076338858859776&0.8820&0.0076&0.7730&0.0135&0.6818&0.0101&0.2943&0.0023& 0.86&10\\
6799537965261994752&0.9000&0.0017&0.9036&0.0092&0.8132&0.0083&0.3267&0.0024& 1.32&24\\
        \hline
    \end{tabular}
    \tablefoot{
     \tablefoottext{o}{High \Ftwo, removed from the Section~\ref{ML-relation} study.} \\
    Full table including the correlations and the orbital parameters will be made available on VizieR. }
    \tablebib{
(1) \cite{2005Obs...125..232T}; (2) \cite{2009A&A...498..949M}; (3) \cite{2018A&A...619A..32B}; (4) \cite{1994AJ....108.1936F}; (5) \cite{2002AJ....124.1132G}; (6) \cite{2009Obs...129..317G}; (7) \cite{2019A&A...626A..31S}; (8) \cite{2015AJ....149...63F}; (9) \cite{1995A&AS..111..255G}; (10) \cite{2018A&A...619A..81H}; (11) \cite{2011Obs...131...17G}; (12) \cite{1990JApA...11..533G}; (13) \cite{2012MNRAS.422...14H}; (14) \cite{2008Obs...128...95G}; (15) \cite{2016MNRAS.458.3272K}; (16) \cite{2013Obs...133....1G}; (17) \cite{2018MNRAS.474..731K}; (18) \cite{1993Obs...113..294G}; (19) \cite{2006RoAJ...16....3I}; (20) \cite{2020MNRAS.496.1355H}; (21) \cite{2007Obs...127..165T}; (22) \cite{2006MNRAS.371.1159G}; (23) \cite{2012Obs...132..356G}; (24) \cite{2019AJ....158..222T}; (25) \cite{2014Obs...134..109G}; (26) \cite{1997MNRAS.284..341M}; (27) \cite{1982A&A...110..241M}
    }
\end{table*}

The uncertainties on the masses of the binary system Gaia DR3 $3954536956780305792$ (HIP~61816) are extremely large, with $\frac{\sigma_{\mathcal{M}_1}}{\mathcal{M}_1} \approx 1$.
These results are therefore unusable. We expect this result to be a consequence of the lack of constrains over the inclination for this system which is $i = 27.5 \pm 12.7 ^{\circ}$. Being compatible with $0$ at less than $3 \sigma$, the masses are much less constrained too, leading to the large uncertainties.\\

The characterization of the binaries from APOGEE is detailed in Table \ref{table-orbit-fitting-apogee}.

A particular case to be considered is Gaia DR3~$839401128363085568$ (LP~129-155).  The results lead to $\mathcal{M}_1 =  1.49 \pm 0.44 \Msun$, $\mathcal{M}_2 = 1.34 \pm 0.37 \Msun$ and $\beta = 0.40 \pm 0.03 $, with $\Ftwo = 2.5$. These parameters make the system an outlier in the mass-luminosity relation. 
Although a MCMC is not adapted to the Thiele-Innes handling, we tested a short MCMC on this star that went to lower mass values, indicating the presence of another solution.
This binary system is the only one of our sample with a low eccentricity consistent with zero at $1 ~\sigma$. The system has then been additionally tested with an eccentricity fixed to $0$, corresponding to a circular orbit. The result obtained fits much better the mass-luminosity relation despite its slightly larger $\Ftwo$, and is the solution kept in Table~\ref{table-orbit-fitting-apogee}.

\begin{table*}[!ht]
    \caption{Solutions from the combination of Gaia NSS astrometric solutions with APOGEE double-lined spectroscopy. $q$ is the mass ratio, $\mathcal{M}_1$ is the mass of the primary, $\mathcal{M}_2$ is the mass of the secondary (in \Msun) and $\beta$ is the flux fraction of the secondary. $\sigma_q$, $\sigma_{\mathcal{M}_1}$, $\sigma_{\mathcal{M}_2}$ and $\sigma_\beta$ are their associated uncertainties. $\Ftwo$ is the goodness-of-fit estimator. }
    \label{table-orbit-fitting-apogee}
    \centering
    \begin{tabular}{lccccccccc}
    \hline
        Gaia DR3 ID & $q$ & $\sigma_q$ & $\mathcal{M}_1$ & $\sigma_{\mathcal{M}_1}$ & $\mathcal{M}_2$ & $\sigma_{\mathcal{M}_2}$ & $\beta$ & $\sigma_\beta$ & $\Ftwo$ \\ \hline \hline
        839401128363085568\tablefootmark{e} & 0.9143 & 0.0993 & 0.6225 & 0.1162 & 0.5692 & 0.0918 & 0.3865 & 0.0266 & 4.03 \\ 
        683525873153063680 & 0.9376 & 0.0496 & 0.5124 & 0.1037 & 0.4804 & 0.0951 & 0.3985 & 0.0167 & 1.72 \\ 
        702393458327135360 & 0.9398 & 0.0252 & 1.4875 & 0.5001 & 1.3980 & 0.4663 & 0.3871 & 0.0186 & 4.53 \\ 
        790545256897189760 & 0.8549 & 0.0846 & 0.4236 & 0.0870 & 0.3621 & 0.0618 & 0.2850 & 0.0263 & 1.20 \\ 
        794359875050204544 & 0.8824 & 0.1294 & 1.1746 & 0.3294 & 1.0365 & 0.2566 & 0.2676 & 0.0360 & 0.99 \\ 
        824315485231630592 & 0.9035 & 0.3744 & 0.9467 & 0.2152 & 0.8554 & 0.2087 & 0.3104 & 0.1036 & 3.92 \\ 
        901170214141646592 & 0.8396 & 0.1134 & 0.9309 & 0.2506 & 0.7815 & 0.1951 & 0.2330 & 0.0404 & 0.46 \\ 
        1267970076306377344 & 0.6633 & 0.0586 & 1.1378 & 0.2014 & 0.7547 & 0.0912 & 0.0703 & 0.0157 & 0.61 \\ 
        1636132061580415488 & 0.7176 & 0.1342 & 1.1046 & 0.4127 & 0.7927 & 0.2574 & 0.1488 & 0.0446 & 3.24 \\ 
        2134829544776833280 & 0.7721 & 0.0680 & 0.8751 & 0.1245 & 0.6757 & 0.0738 & 0.1505 & 0.0251 & 0.73 \\ 
        2705239237909520128 & 0.5293 & 0.1035 & 0.2336 & 0.0683 & 0.1236 & 0.0326 & 0.1993 & 0.0427 & 0.53 \\ 
        3847995791877023104 & 0.8387 & 0.0788 & 1.1446 & 0.2029 & 0.9600 & 0.1550 & 0.3167 & 0.0244 & 1.05 \\ 
        5285071954833306368\tablefootmark{o} & 0.7905 & 0.1019 & 0.4609 & 0.1058 & 0.3643 & 0.0853 & 0.1499 & 0.0416 & 1.58 \\ \hline
    \end{tabular}
    \tablefoot{
    \tablefoottext{e}{particular case for which the eccentricity has been fixed to 0.}
    \tablefoottext{o}{Mass-luminosity outlier, removed from the Section~\ref{ML-relation} study.} \\
    Full table including the correlations and the orbital parameters will be made available on VizieR. }
\end{table*}

Since Gaia provides the $G$ magnitude of the binary system, the individual absolute magnitude in the $G$-band can be deduced for each star using $A_G$ the extinction in the $G$ band derived from the 3D extinction map of \cite{Lallement22} and using the Gaia DR3 extinction law\footnote{https://www.cosmos.esa.int/web/Gaia/edr3-extinction-law}:
\begin{equation} \label{absolute-magnitude-G-primary}
M_{G_1} = G + 2.5 \log_{10}\left(\frac{1}{1-\beta}\right) + 5 + 5 \log_{10} \left(\frac{\varpi}{1000}\right) - A_G
\end{equation}
\begin{equation} \label{absolute-magnitude-G-secondary}
M_{G_2} = G + 2.5 \log_{10}\left(\frac{1}{\beta}\right) + 5 + 5 \log_{10} \left(\frac{\varpi}{1000}\right) - A_G
\end{equation}
where $\varpi$ is the parallax in mas.

To estimate the uncertainties of those absolute magnitude (see Appendix~\ref{annex-error-jacobian-calculation}), a 10\% relative error on the extinction with a minimum error of 0.01 mag is assumed and a 0.01~mag error is quadratically added to the $G$ formal magnitude errors.
The extinction term $A_G$ remain negligible for 90\% of our sample, with a maximum $A_G$ of 0.03 for APOGEE and 0.15 for SB9.

\begin{figure}[h!]
    \includegraphics[scale=0.5]{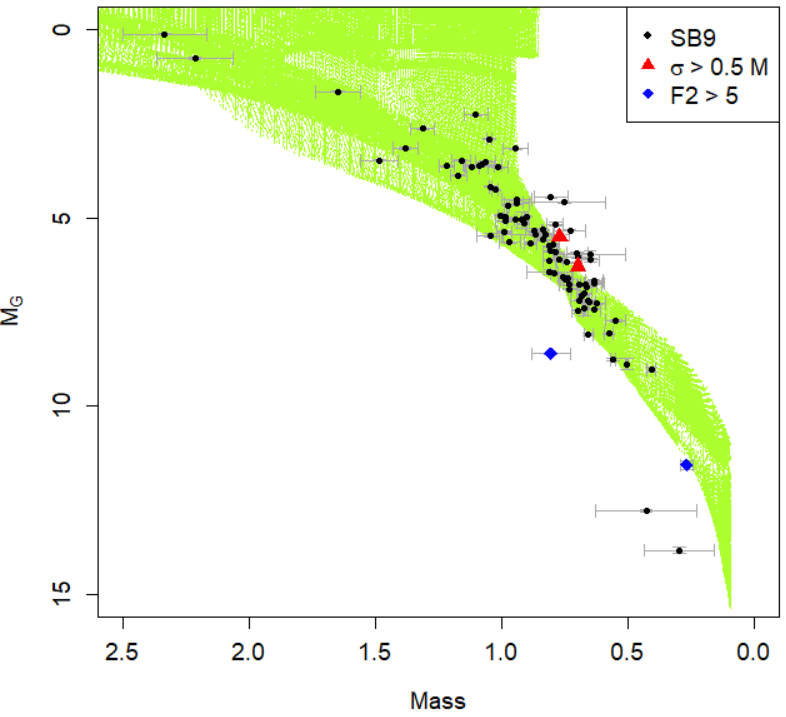}
    \caption{Mass-luminosity diagram of the characterized stars from the combination of Gaia with SB9. The error bars at $1\sigma$ are given in grey. The absolute magnitude of the individual stars in the $G$-band $M_G$ is plotted as a function of the star mass (in \Msun). The black dots represent the stars, with the associated error bars at $1 \sigma$ in grey. The red triangles represent Gaia DR3 $3954536956780305792$, for which the uncertainties are really large and not represented here. The blue diamonds represent Gaia DR3 $3427930123268526720$ for which $\Ftwo > 5$. They are overplotted on the solar-metallicity PARSEC isochrones (in green).
    }
    \label{ml-diag-error-sb9}
\end{figure}

\begin{figure}[h!]
    \includegraphics[scale=0.5]{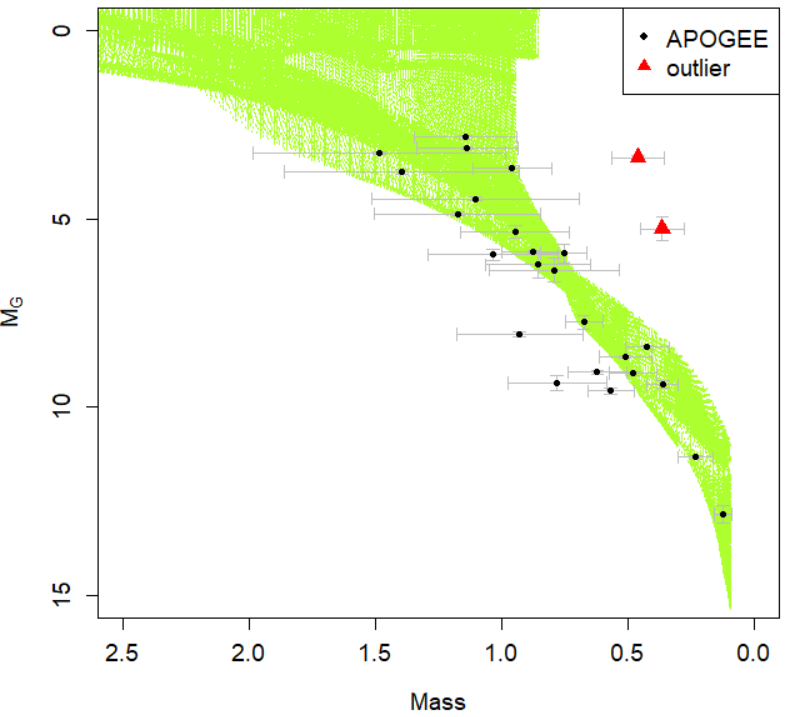}
    \caption{Mass-luminosity diagram of the characterized stars from the combination of Gaia with APOGEE. The error bars at $1~\sigma$ are given in grey. The absolute magnitude of the individual stars in the $G$-band $M_G$ is plotted as a function of the star mass (in \Msun). The black dots represent the stars, with the associated error bars at $1 \sigma$ in grey. The red triangles represent the outlier Gaia DR3 ID $5285071954833306368$. They are overplotted on the isochrones (in green).}
    \label{ml-diag-error-apogee}
\end{figure}

Figure \ref{ml-diag-error-sb9} for SB9 and Figure \ref{ml-diag-error-apogee} for APOGEE give the position of all the individual stars we have characterized in the mass-luminosity diagram.
They are overplotted on top of the PARSEC solar-metallicity isochrones \citep[PAdova and TRieste Stellar Evolution Code,][]{PARSEC}. 

Almost all the stars are compatible with the isochrones at $3 \sigma$. 
For SB9, one star can be considered as an outlier, Gaia DR3 $3427930123268526720$ (GJ~220), which is the only system of the SB9 sample with $\Ftwo > 5$. 
For APOGEE, one main outlier exists, Gaia DR3 $5285071954833306368$ (HD~50199), still compatible with the isochrones at less than $5 \sigma$. Nothing specific has been found about this system to justify its surprising position in the diagram.

11 systems have a goodness-of-fit of their Gaia DR3 astrometric solution higher than 10, but none are outliers in the mass-luminosity relation nor in the \Ftwo of the combined fit nor in the relation between the flux fraction and the mass ratio. We therefore decided to keep those systems in our sample for the Section~\ref{ML-relation} mass-luminosity relation study.

\subsection{Comparison with the direct calculation method for SB9}

As a sanity check, we compare the masses obtained with the orbit fitting process (Table~\ref{table-orbit-fitting-sb9}) to those obtained through direct calculation, that is using directly the orbital parameters provided by SB9 to derive the mass functions without going back to the raw data, as detailed in Annex~\ref{annex-direct-calculation-sb9} (Table~\ref{table-direct-calculation-sb9}). 

Going back to the raw spectroscopic data 
allows to take into account the correlations between the spectroscopic parameters and then have a better estimation of the orbital parameters and their uncertainties.
Moreover, a correction process is applied to the uncertainties of the radial velocity epochs making them more realistic.

\begin{figure}[h!]
    \includegraphics[scale=0.5]{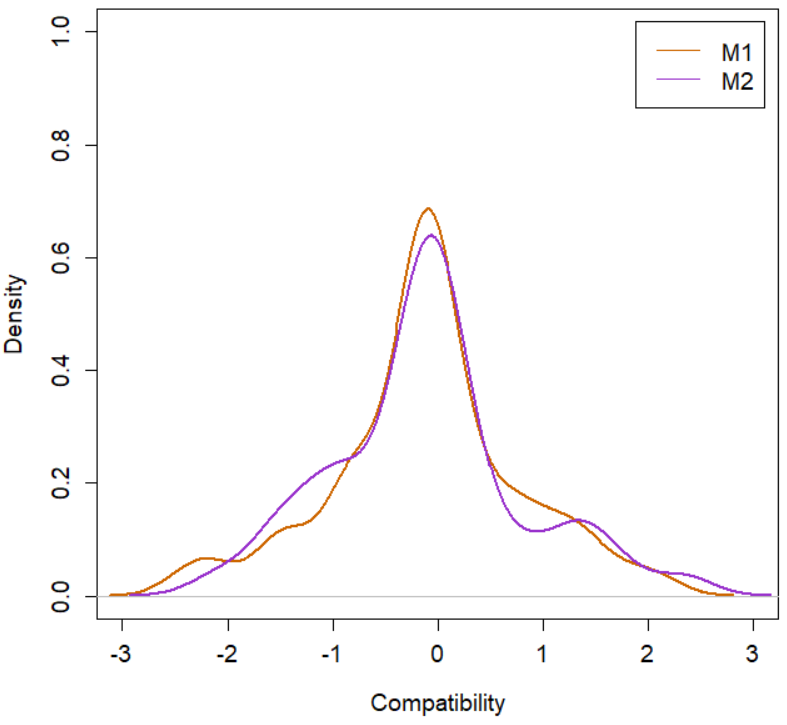}
    \caption{Compatibility density of the masses obtained by direct calculation and by orbit fitting for SB9 binaries. The compatibility is in $\sigma$. It is given in orange for primary masses and in purple for secondary masses.}
    \label{mass-comparison-relative-diff}
\end{figure}

Figure \ref{mass-comparison-relative-diff} presents the distribution of the compatibility (in $\sigma$) between the masses obtained for the orbit fitting process and the direct calculation. The compatibility is defined as:
\begin{equation} \label{eq-relative-difference}
    {\rm compatibility} = \frac{\mathcal{M}_{OF} - \mathcal{M}_{DC}}{\max(\sigma_{OF}, \sigma_{DC})}
\end{equation}
where $\mathcal{M}_{OF}$ is the mass obtained through orbit fitting and $\mathcal{M}_{DC}$ the mass obtained through direct calculation.

It shows that as expected the results are nicely compatible, but with a difference that is not fully negligible in some cases.

\begin{figure}[h!]
   \includegraphics[scale=0.5]{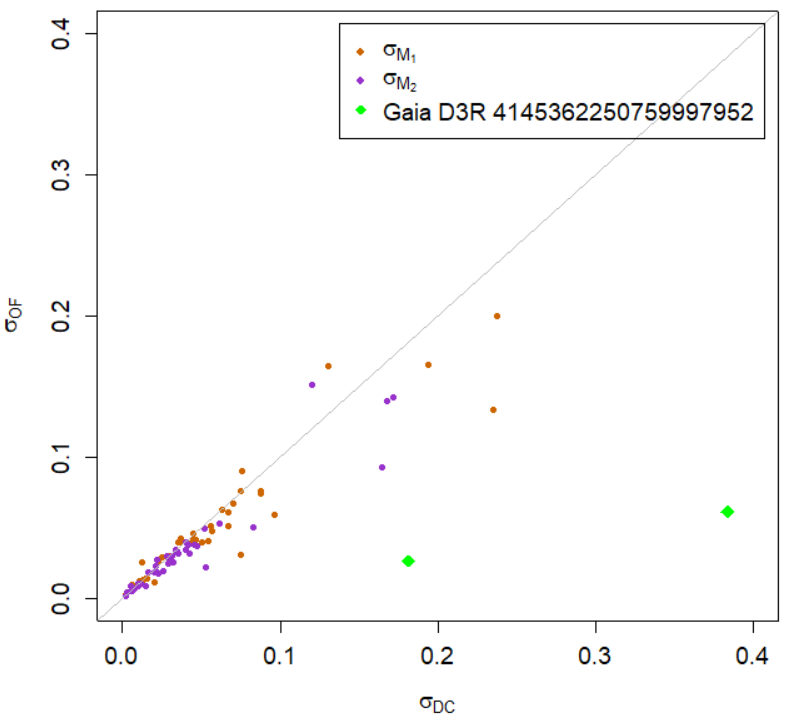}
    \caption{Comparison of the uncertainties on the masses obtained by direct calculation ($\sigma_{DC}$) or by orbit fitting ($\sigma_{OF}$) for binaries from SB9. The primaries are given in orange, and secondaries in purple. The outlier star system Gaia DR3 $4145362250759997952$ is represented in green for its primary and secondary. The grey line is the identity line $y = x$.}
    \label{error-comparison}
\end{figure}

Figure \ref{error-comparison} provides the uncertainties over the masses obtained with one method with respect to the other.
The majority of the mass uncertainties are under the identity line, meaning that the uncertainties coming from the direct calculation process are generally higher than the ones from  orbit fitting. It confirms 
that the uncertainties over the orbital parameters are often reduced when accounting for the correlations between them.

A few points have slightly bigger uncertainties through the orbit fitting process. This is the result of the correction process applied to the uncertainties (see Eq.~\ref{equation-correction-error}).

One noteworthy case is Gaia DR3 $4145362250759997952$ (HD~163336~B), for which the difference between the uncertainties is quite large. The radial velocity data contains only 3 epochs for the secondary. Thus, we can expect a strong correlation between the SB9 orbital parameters. We performed a MCMC over the raw spectroscopic data only and confirm 
that strong correlations appear and that the distribution is strongly asymmetric, explaining the strong improvement obtained by including the knowledge of the Gaia orbit in the spectroscopic fit. 

\subsection{Reference comparison}

Three star systems from our SB9 sample are identified as SB2 in the Gaia DR3 catalogue with direct masses derived by \cite{DR3NSS}. While our mass estimates are consistent with Gaia DR3 1067685718250692352 (HIP~45794) and Gaia DR3 2035577729682322176 (HIP~97640), Gaia DR3 595390807776621824 HIP~42418) is a 5$\sigma$ outlier. It has a photocentre semi-major axis $a_0$ of 3.4 mas while the other two stars have a smaller $a_0\sim0.8$ mas, suggesting that the astrometric motion impacted the spectroscopic measure, as suggested by \cite{DR3Validation}. This is due to the fact that the expected position of the spectra on the Gaia Radial Velocity Spectrometer (RVS) detectors is predicted by the standard 5-parameter astrometric motion instead of the epoch astrometric one which would not be precise enough. 

\begin{figure}[h!]
    \includegraphics[scale=0.5]{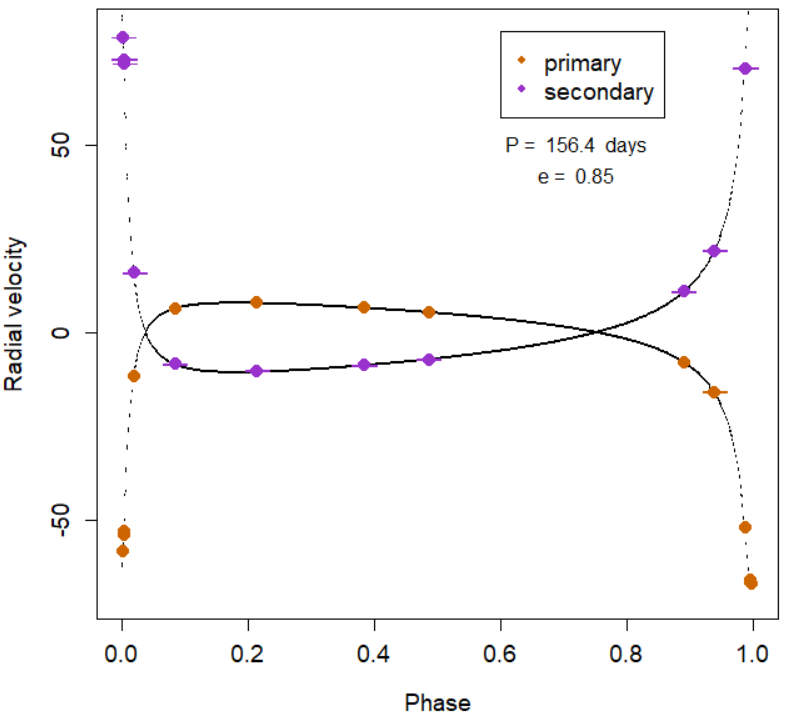}
    \caption{Radial velocity fit for the binary system Gaia DR3 $3283823387685219328$ (HIP~20601) from SB9. The radial velocity (km/s) is plotted as a function of the phase. The orange dots correspond to the radial velocity epochs of the primary, and the purple dots give the radial velocity epochs for the secondary. The black curves represent the corresponding fits obtained combining the SB9 epoch radial velocity with the Gaia DR3 NSS astrometric orbital solution with BINARYS.
    }
    \label{rv-sb9}
\end{figure}

Masses were obtained combining a visual orbit with an SB2 orbit for Gaia DR3 $2129771310248902016$ (HIP 95575) by \cite{2020MNRAS.492.2709P} ($\mathcal{M}_1 = 0.670 \pm 0.069$, $\mathcal{M}_2 =  0.602 \pm 0.061$, compatible with our results within $1\sigma$), for Gaia DR3 $2067948245320365184$ (HIP 101382) by \cite{2018MNRAS.474..731K} ($\mathcal{M}_1 = 0.8420 \pm 0.0014$, $\mathcal{M}_2 =  0.66201 \pm 0.00076$, compatible with our results within $2\sigma$) and for Gaia DR3 $3283823387685219328$ (HIP 20601) by \cite{2020MNRAS.496.1355H} ($\mathcal{M}_1 = 0.9798 \pm 0.0019 \Msun$, $\mathcal{M}_2 = 0.72697 \pm 0.00094 \Msun$, compatible with our results within $1\sigma$). \cite{2020MNRAS.496.1355H} combined the raw relative astrometry and spectroscopic data. They obtained a parallax at 
4.4$\sigma$ from the Gaia NSS one. 
We tested a combined fit with the radial velocity, interferometry from \cite{2020MNRAS.496.1355H} 
and Gaia NSS solution and obtained a goodness of fit of $\Ftwo=1.5$, a parallax of $\varpi=16.573\pm0.017$ and masses $\mathcal{M}_1 = 0.9816 \pm 0.0014  \Msun$ and $\mathcal{M}_2 = 0.72808 \pm 0.00076 \Msun$. This new parallax is at 3.4$\sigma$ from the \cite{2020MNRAS.496.1355H} one and reduced to 2.6$\sigma$ if we take into account the Gaia DR3 parallax zero point \citep{LindegrenEDR3bias}, highlighting that SB2 stars with direct imaging will be excellent test cases for Gaia DR4 epoch data validation. 
The orbit fit for this binary is given on the Figure \ref{rv-sb9}.

\begin{figure}[h!]
    \includegraphics[scale=0.5]{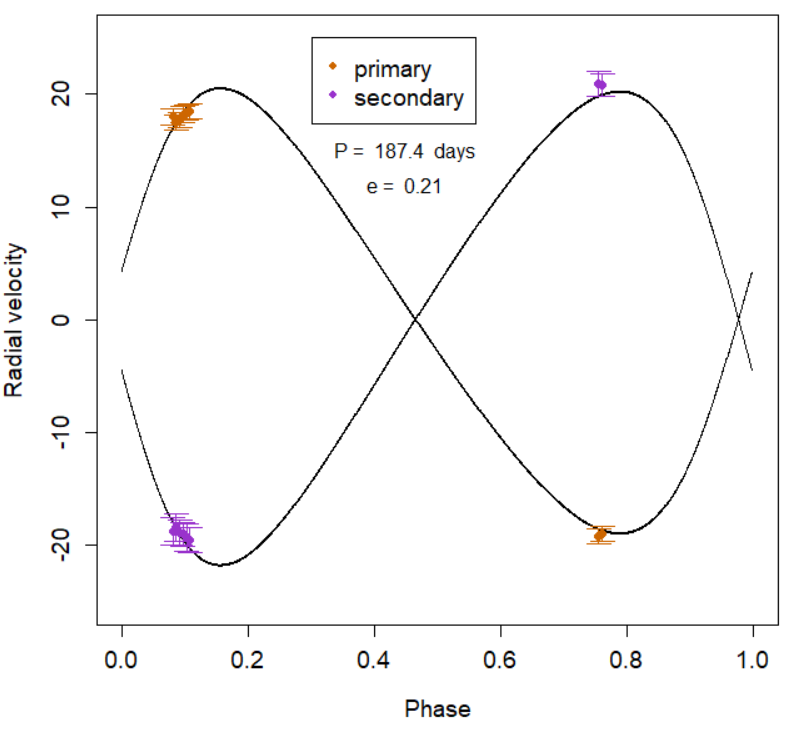}
    \caption{Radial velocity fit for the binary system Gaia DR3 $702393458327135360$ (HD~80234) from APOGEE. The radial velocity (km/s) is plotted as a function of the phase. The orange dots correspond to the radial velocity epochs of the primary, and the purple dots give the radial velocity epochs for the secondary. The black curves represent the corresponding fits obtained combining the APOGEE epoch radial velocity with the Gaia DR3 NSS astrometric orbital solution with BINARYS.
    }
    \label{rv-apogee}
\end{figure}

For APOGEE, the only star system for which masses have been obtained in the literature is the binary Gaia DR3 $702393458327135360$ which has been discussed in Section \ref{data-presentation-SB2}. 
This star has been solved through spectroscopy only \citep{APOGEESB2} and then combined with Gaia astrometry through a direct calculation process by \cite{DR3NSS} to obtain $\mathcal{M}_1 = 1.14 \pm 0.38 \Msun$, $\mathcal{M}_2 = 1.06 \pm 0.35 \Msun$ and $F_2/F_1 = 0.567 \pm 0.071$, 
fully compatible with our results. The uncertainties reported here are larger than the one of \cite{DR3NSS}. This may be due to a slight discrepancy between the orbital parameters of APOGEE and Gaia, specifically on the eccentricities which are at 4$\sigma$ from each other. This discrepancy leads to a high $\Ftwo$ of $4.6$ in our solution and to higher uncertainties than what is obtained by a direct calculation.
The orbit fit for the binary Gaia DR3 $702393458327135360$ (HD~80234) is given on the Figure \ref{rv-apogee}. 
The difference of observation time clearly appears on the Figures \ref{rv-sb9} and \ref{rv-apogee}, where the lack of radial velocity epochs for APOGEE is rather obvious. 

\section{The Mass-Luminosity relation} \label{ML-relation}

The mass calculations presented above could enable to perform an empirical fit of the mass-luminosity relation in the G-band using the Gaia photometry. However, the masses calculated 
do not provide satisfying constrains in the interesting region of the relation for low-mass stars ($\mathcal{M} < 0.7 \Msun$).
To fill in this part of the H-R diagram, we looked for low mass stars resolved by Gaia and direct imaging data from the literature, following the work of \cite{BINARYS} on HIP 88745. 

\subsection{Low-mass systems resolved by Gaia with direct imaging data \label{sec:resolvedstars}}

We found three star systems spatially resolved studied with direct imaging in \cite{MLrelation} which also have Hipparcos \citep{HIP2} transit data (TD) and Gaia resolved observations consistent with the direct imaging data:  \object{Gl~330}, \object{Gl~860} and \object{Gl~277}. 
\object{Gl~568} is not in the \cite{MLrelation} sample but has direct imaging data from \cite{1989AJ.....97..510M} and \cite{2018AJ....155..215M} and has been added to our sample. Those four stars are analysed by BINARYS by combining the direct imaging data with Hipparcos TD and Gaia astrometric parameters of both components following the methodology detailed in \cite{BINARYS}. All Gaia solutions have a 5-parameter solution except Gl~330 for which the secondary component is a 2-parameter solution and Gl~860 for which both components are 2-parameter solutions only. 
One Hipparcos TD outlier at 5$\upsigma$ had to be removed for both Gl~568 and Gl~227. 

Two stars from the \cite{MLrelation} sample, \object{Gl~65} and \object{Gl~473}, 
are resolved by Gaia with separations consistent with the visual orbit,
but no Hipparcos data exist for them.
However those two stars have literature mass fraction $B=\frac{\mathcal{M}_2}{\mathcal{M}_1 + \mathcal{M}_2}$: for Gl~65 $B=0.494\pm0.04$ from \cite{Geyer88} and for Gl~473 $B = 0.477\pm0.008$ from \cite{Torres99}. We incorporated this information within BINARYS for those stars.

As our sample contains very nearby stars, we added the perspective acceleration terms in BINARYS following the description detailed in \cite{HIP} and \cite{NSSastro}. We first compute the radial proper motion, that is the relative change in distance per year, $\mu_r = V_r \varpi / A_Z$ in yr$^{-1}$ with $A_Z=9.7779222 \times 10^8$ mas\,yr\,km\,s$^{-1}$. The perspective acceleration changes the along-scan abscissa $\nu$ (in mas) by adding :
\begin{equation}
    \Delta \nu = -\mu_r \Delta T \left( \frac{\partial \nu}{\partial \varpi} \varpi + 
    \frac{\partial \nu}{\partial \mu_{\alpha^*}} \mu_{\alpha^*} + 
    \frac{\partial \nu}{\partial \mu_{\delta}} \mu_{\delta} \right)
\end{equation}
with $\Delta T$ the epoch in years relative to the reference epoch for the astrometric parameters (that is 1991.25 for Hipparcos and 2016.0 for Gaia DR3). This $\Delta \nu$ is subtracted to the Hipparcos abscissa residuals and added to the Gaia simulated abscissa. However one has to take into account that the perspective acceleration has been taken into account for DR3 for stars with a Gaia DR2 radial velocity or in the table of nearby Hipparcos stars with radial velocity used for DR2\footnote{
\url{https://gea.esac.esa.int/archive/documentation/GDR2/Data\_processing/chap\_cu3ast/sec\_cu3ast\_cali/ssec\_cu3ast\_cali\_source.html\#Ch3.T3}}. 
Here only the radial velocity for the A component of Gl~860 was applied for the DR3 processing, using the same $V_r=-33.94$ \kms as we used. For Gl~65 we used $V_r=39.04$ \kms \citep{Kervella16} while for Gl~473 all literature $V_r$ are consistent with zero. We checked that taking into account the perspective acceleration for our stars only change marginally the $\chi^2$. 

The results obtained for these six stars are given in Table \ref{table-results-stars-mann}. 
Our mass estimates are consistent with \cite{Kervella16}, \cite{ML_delfosse} and \cite{ML_benedict} for Gl~65, but we all use the same literature mass fraction. For Gl~473, using again the same literature mass fraction, our masses are consistent with \cite{ML_delfosse} but are at 2.5$\sigma$ from the values of \cite{ML_benedict} which is driven by the difference with the RECONS parallax they used. Our mass estimate for Gl~860 is consistent with \cite{ML_delfosse}. We derive the dynamical masses for the first time, to our knowledge, of the components of Gl~277, Gl~330 and Gl~568.

\begin{table*}[!ht]
    \caption{Solutions for star systems resolved by Gaia with direct imaging data and either Hipparcos Transit Data or literature mass fraction. The masses of the primary $\mathcal{M}_1$ and secondary $\mathcal{M}_2$ are in \Msun, the parallax $\varpi$ in mas, $\rho_{\mathcal{M}_1,\mathcal{M}_2}$ is the correlation between the mass estimates and \Ftwo is the goodness of fit. }
    \label{table-results-stars-mann}
    \centering
    \begin{tabular}{lccccrcrc}
    \hline
         Name & $\mathcal{M}_1$ & $\sigma_{\mathcal{M}_1}$ & $\mathcal{M}_2$ & $\sigma_{\mathcal{M}_2}$ & $\varpi$ & $\sigma_{\varpi}$ & $\rho_{\mathcal{M}_1,\mathcal{M}_2}$ & $\Ftwo$   \\ \hline \hline
Gl277&0.5276&0.0046&0.2069&0.0031&83.465&0.052&0.77&3.93\\
Gl330&0.4969&0.0247&0.3184&0.0159&61.513&0.107&0.98&3.37\\
Gl568&0.3106&0.0067&0.2129&0.0047&87.439&0.069&0.95&1.53\\
Gl860&0.2915&0.0062&0.1789&0.0046&248.354&1.560&0.39&3.49\\
Gl65\tablefootmark{*}&0.1224&0.0012&0.1195&0.0012&371.396&0.453&$-0.28$&1.19\\
Gl473\tablefootmark{*}&0.1379&0.0023&0.1258&0.0022&227.041&0.389&$-0.76$&1.36\\
\hline
\end{tabular}
\tablefoot{\tablefoottext{*}{Solution using a literature mass fraction.} Full table including Gaia and Hipparcos IDs and full orbital solution will be made available on VizieR. }
\end{table*}

\subsection{Fitting the Mass-Luminosity relation \label{sec:mlfit}}

To fit the mass-luminosity relation, we implemented a TMB function that allows to take into account the uncertainties on both the mass and the magnitude, and more importantly, the correlations between the parameters of the two components of the same system (the calculation of the covariance is detailed in Annex~\ref{annex-error-jacobian-calculation}).
The true magnitudes of the stars are used as a random parameter: they are marginalized, that is integrated out of the likelihood. The initial value of TMB is provided by a classical polynomial fit (\texttt{R lm} function). We selected only components with $M_G>5$ to be used as input for the fit as for fainter magnitudes the age dependency is well known to be too large. 

We tested several degree for the polynomial and fitting the logarithm of the mass instead of the mass itself and used the Bayesian Information Criterion (BIC) to compare the models. The BIC favoured a polynomial of degree 4 fitting the log of the mass. The coefficients are given in Table \ref{table-fit-tmb-results}. The fit uncertainties have been estimated through a bootstrap, leading to uncertainties smaller than 0.015\Msun for magnitudes higher than $M_G>6$ corresponding to masses $\mathcal{M}<0.77$~\Msun.

\begin{figure}[h!]
    \centering
    \includegraphics[width=9cm]{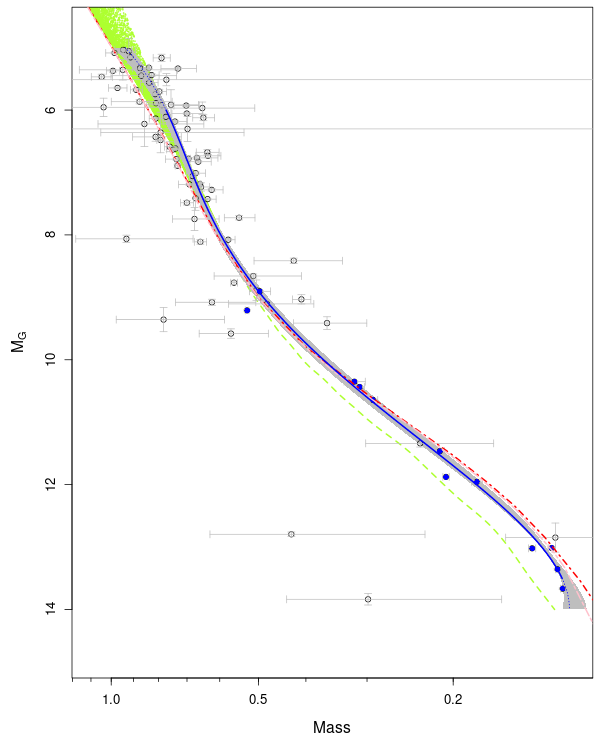}
    \caption{Mass-luminosity relation fitted in this work (in blue), with areas outside the $6<M_G<13.5$ range as dotted line. The grey area corresponds to the 1$\sigma$ bootstrap interval. The stars from Table~\ref{table-results-stars-mann} (resolved Gaia stars with direct imaging data) are in blue and stars from Tables~\ref{table-orbit-fitting-sb9} and \ref{table-orbit-fitting-apogee} (SB2 data combined with Gaia NSS astrometric solution) are in grey. 
    Dashed green lines are the solar metallicity PARSEC isochrones for main sequence stars (isochrones with label=1). 
    The dotdashed red line and the twodashed pink line are the solar metallicity isochrone of \cite{Baraffe15} and BASTI respectively; they show a deviation only at masses lower than $\sim$0.12\Msun.}.
    \label{ML-fit-tmb}
\end{figure}

The fit is displayed on Fig.~\ref{ML-fit-tmb} together with the PARSEC, the \cite{Baraffe15}\footnote{http://perso.ens-lyon.fr/isabelle.baraffe/BHAC15dir/} and the BASTI \citep[A BAg of Stellar Tracks and Isochrones,][]{BASTI}\footnote{http://basti-iac.oa-teramo.inaf.it/} isochrones. 
The age dependency of the mass-luminosity relation starts to be significant for all isochrones at  $\gtrsim 0.6$~\Msun and our fit follows the oldest isochrones.
For masses $<0.5\Msun$ our empirical relation indicate lower masses for a given luminosity than the PARSEC isochrones. Our results are consistent with the \cite{Baraffe15} isochrones except in the low mass region where we find slightly higher masses for a given luminosity.

\begin{table}[!h]
    \centering
    \caption{Coefficient values of the polynomial fit $ \log_{10}(\mathcal{M}) = C_0 + C_1 M_G + C_2 M_G^2 + C_3 M_G^3 + C_4 M_G^4 $, to be used within
    $6<M_G<13.5$ which corresponds to the
    $0.12<\mathcal{M}<0.77$ mass range. The coefficients are provided for the best fit as well as the upper and lower curves at 1$\sigma$ obtained through a bootstrap. }
    \label{table-fit-tmb-results}
    \begin{tabular}{cS[table-format=-1.7]S[table-format=-1.7]S[table-format=-1.7]}
    \hline
        & {best fit} & {upper} & {lower} \\ 
        \hline
        \hline
    C$_0$&3.129&3.798&2.534\\
    C$_1$&-1.5406&-1.8377&-1.2672\\
    C$_2$&0.27513&0.32373&0.22954\\
    C$_3$&-0.021661&-0.025067&-0.018413\\
    C$_4$&0.0005991&0.0006855&0.0005148\\
        \hline
    \end{tabular}
\end{table}

\section{Conclusion} \label{conclusion}

We have estimated the masses of binary systems by combining the astrometric orbits from Gaia DR3 with spectroscopy, 43 star systems from SB9 and 13 from APOGEE. While the spectroscopic orbit was already known for the SB9 stars, it was the case for only one APOGEE star. We tested on SB9 the difference between a direct calculation of the masses using the orbital parameters and a combined fit using the raw radial velocity measures, the later estimating better the parameters, their uncertainties and their correlations. 
We also estimated the masses of 6 stars resolved by Gaia DR3 with literature direct imaging and either Hipparcos data or literature mass fraction. Three of those stars have dynamical masses derived for the first time. 

The BINARYS tool have been used to perform the combined fits. BINARYS was extended for this study to handle Gaia DR3 NSS solutions and perspective acceleration within the Hipparcos and Gaia observation time.

Using the derived masses and the $G$ magnitudes we derived a first empirical mass-luminosity relation in the $G$ band taking into account all the correlations between the component masses and magnitudes. This empirical relation is found to be in better agreement with the \cite{Baraffe15} isochrones than with the PARSEC ones. 

We expect Gaia DR4 to significantly increase the sample of stars that could be used in such a study. Moreover Gaia DR4 will provide access to the epoch astrometry. It will enable to make a full combined fit on the raw data of both spectroscopic or direct imaging data with Gaia astrometry. It will also enable to dig into systems with a too low astrometric signal to have a full Gaia DR4 orbital solution but with a good spectroscopic or direct imaging one. We can therefore expect a much more in depth study of the mass-luminosity relation with Gaia DR4, in particular a study of the metallicity dependency should be conducted with a larger sample.

\begin{acknowledgements}
We thank J.B. Le Bouquin for helping debugging our test on HIP 20601 and X. Delfosse for providing lists of close-by M-dwarfs to dig into. We thank S. Cassisi for his prompt feedback on the BASTI isochrones.
T.M. is granted by the BELSPO Belgian federal research program FED-tWIN under the research profile Prf-2020-033\_BISTRO.
This work has made use of data from the European Space Agency (ESA) space mission Gaia (\url{https://www.cosmos.esa.int/gaia}), processed by the Gaia Data Processing and Analysis Consortium (DPAC). Funding for the DPAC is provided by national institutions, in particular the institutions participating in the Gaia MultiLateral Agreement. 
This research has made use of the SIMBAD database, operated at CDS, Strasbourg, France.
\end{acknowledgements}

\bibliographystyle{aa}
\bibliography{bibfile}

\appendix

\section{The direct calculation process for SB9} \label{annex-direct-calculation-sb9}

 Since the orbital parameters are both given by SB9 and Gaia, the combination of astrometry with spectroscopy can be done easily. The features of the star systems are directly computed from the astrometric and spectroscopic orbital solutions, using commonly used formulae.
 This direct-calculation method is approximately the same as applied for mass calculations by the Gaia collaboration \citep{DR3NSS}. 

From the semi-amplitudes given by spectroscopy, the mass ratio is
\begin{equation} \label{eq-mass-ratio}
     q = \frac{\mathcal{M}_2}{\mathcal{M}_1} = \frac{K_1}{K_2}.
\end{equation}

The dynamical mass of the secondary is 
\begin{equation} \label{eq-secondary-mass}
    \mathcal{M}_2 = 1.0385 \times 10^{-7} K_1^3 P (1-e^2)^{3/2} \left( 1+\frac{1}{q} \right)^2 \frac{1}{\sin^3{i}},
\end{equation}
with $P$ the period in days, $K_1$ in km.s$^{-1}$
and the inclination $i$ given by the Gaia astrometry.
The period and the eccentricity $e$ used are the weighted mean of the Gaia and SB9 values. 

The primary mass is then deduced from the mass ratio and the secondary mass.

The flux fraction of the secondary $\beta = \frac{F_2}{F_1+F_2}$ is given through the relation between the individual stars orbits and the photocentre orbit provided by astrometry:
\begin{equation} \label{eq-orbit-relation}
    a_0 = (B - \beta)~ a_{21}
\end{equation}

with $B$ the mass fraction of the secondary 
\begin{equation}\label{eq-mass-fraction}
    B = \frac{\mathcal{M}_2}{\mathcal{M}_1+\mathcal{M}_2} = \frac{q}{1+q},
\end{equation}
$a_{21}$ the semi-major axis of the relative orbit between the primary and the secondary 
\begin{equation}\label{eq-relative-orbit}
    a_{21} = a_1+a_2 = a_1 \left( 1 + \frac{1}{q} \right),
\end{equation}
and $a_0$ the semi-major axis of the the photocentre orbit.

Here, $a_{1/2}$ are the individual semi-major axis of the primary/secondary respectively. They can be deduced from the semi-amplitudes:
\begin{equation} \label{eq-individual-orbit}
    a_{1, 2} = \frac{1}{10879} K_{1, 2} P \frac{\sqrt{1-e^2}}{\sin{i}}
\end{equation}
with the semi-major axis in au and the period in days.

The masses $\mathcal{M}_1$, $\mathcal{M}_2$, and the flux fraction $\beta$ are then obtained through Eq.~\ref{primary-mass-jacobian}, \ref{secondary-mass-jacobian}, \ref{eq-def-beta} respectively.

Astrometry is needed here to determine the photocentre orbit $a_0$ but also to convert it from mas to au using the parallax $\varpi$. \\

The errors are estimated through a Monte-Carlo (with $n_{MC} = 10^3$) assuming all the parameter distributions to be Gaussian-like.
The Gaia parameters are generated simultaneously according to the covariance matrix, except the photocentre semi-major axis $a_0$. Its error has been shown by \cite{DR3Validation} to be over-estimated by a Monte-Carlo approach and the local linear approximations of \cite{NSSastro} have been used for it.
In order to take into account potential asymmetry in final parameters distributions, the lower and upper confidence levels at respectively 16\% and 84\% are determined. 

The results of this direct calculation are given in Table \ref{table-direct-calculation-sb9}.

\begin{table*}[!ht]
    \caption{Results from direct calculation : SB9}
    \label{table-direct-calculation-sb9}
    \centering
    \begin{tabular}{ccccccccccccc}
    \hline
         Gaia DR3 ID & $\mathcal{M}_1$ & $\sigma_{\mathcal{M}_1}$ & $\mathcal{M}_{1_{low}}$ & $\mathcal{M}_{1_{up}}$ & $\mathcal{M}_{2}$ & $\sigma_{\mathcal{M}_2}$ & $\mathcal{M}_{2_{low}}$ & $\mathcal{M}_{2_{up}}$ & $\beta$ & $\sigma_{\beta}$ & $\beta_{low}$ & $\beta_{up}$ \\ \hline \hline
        48197783694869760 & 0.994 & 0.038 & 0.954 & 1.031 & 0.679 & 0.022 & 0.656 & 0.700 & 0.101 & 0.001 & 0.097 & 0.105 \\ 
        69883417170175488 & 0.895 & 0.024 & 0.873 & 0.920 & 0.843 & 0.021 & 0.822 & 0.863 & 0.399 & 0.003 & 0.394 & 0.404 \\ 
        308256610357824640 & 0.440 & 0.341 & 0.168 & 0.478 & 0.312 & 0.241 & 0.119 & 0.338 & 0.283 & 0.010 & 0.233 & 0.288 \\ 
        478996438146017280 & 0.778 & 0.022 & 0.753 & 0.797 & 0.717 & 0.020 & 0.694 & 0.734 & 0.385 & 0.002 & 0.382 & 0.388 \\ 
        544027809281308544 & 0.954 & 0.059 & 0.897 & 1.018 & 0.796 & 0.036 & 0.761 & 0.833 & 0.137 & 0.001 & 0.131 & 0.144 \\ 
        595390807776621824 & 0.788 & 0.011 & 0.777 & 0.799 & 0.695 & 0.007 & 0.687 & 0.702 & 0.309 & 0.001 & 0.307 & 0.312 \\ 
        660622010858088320 & 0.715 & 0.065 & 0.656 & 0.783 & 0.626 & 0.038 & 0.588 & 0.665 & 0.227 & 0.003 & 0.217 & 0.237 \\ 
        827608625636174720 & 0.776 & 0.051 & 0.726 & 0.823 & 0.730 & 0.049 & 0.682 & 0.775 & 0.408 & 0.003 & 0.404 & 0.412 \\ 
        882872210352301568 & 1.007 & 0.011 & 0.996 & 1.018 & 0.506 & 0.003 & 0.503 & 0.509 & 0.025 & 0.001 & 0.021 & 0.029 \\ 
        1067685718250692352 & 1.035 & 0.013 & 1.022 & 1.048 & 0.919 & 0.012 & 0.908 & 0.931 & 0.300 & 0.002 & 0.294 & 0.305 \\ 
        1074883087005896320 & 0.641 & 0.037 & 0.608 & 0.680 & 0.602 & 0.034 & 0.572 & 0.638 & 0.396 & 0.005 & 0.389 & 0.403 \\ 
        1324699172583973248 & 1.309 & 0.056 & 1.254 & 1.365 & 1.154 & 0.044 & 1.109 & 1.198 & 0.315 & 0.002 & 0.311 & 0.319 \\ 
        1441993625629660800 & 1.116 & 0.046 & 1.062 & 1.151 & 0.932 & 0.028 & 0.897 & 0.951 & 0.092 & 0.001 & 0.084 & 0.098 \\ 
        1480959875337657088 & 0.864 & 0.257 & 0.655 & 1.122 & 0.811 & 0.169 & 0.664 & 0.978 & 0.291 & 0.014 & 0.258 & 0.324 \\ 
        1517219363639758976 & 1.114 & 0.053 & 1.066 & 1.170 & 0.913 & 0.042 & 0.875 & 0.958 & 0.163 & 0.001 & 0.157 & 0.170 \\ 
        1517927895803742080 & 0.584 & 0.013 & 0.571 & 0.597 & 0.549 & 0.017 & 0.533 & 0.566 & 0.361 & 0.002 & 0.357 & 0.365 \\ 
        1528045017687961856 & 1.125 & 0.030 & 1.095 & 1.154 & 0.764 & 0.019 & 0.744 & 0.783 & 0.068 & 0.001 & 0.064 & 0.071 \\ 
        1615450866336763904 & 1.000 & 0.045 & 0.956 & 1.044 & 0.787 & 0.031 & 0.757 & 0.816 & 0.131 & 0.001 & 0.125 & 0.136 \\ 
        1918953867019478144 & 1.040 & 0.006 & 1.036 & 1.047 & 0.969 & 0.005 & 0.964 & 0.974 & 0.390 & 0.001 & 0.388 & 0.392 \\ 
        2012218158438964224 & 2.072 & 0.129 & 1.957 & 2.207 & 1.952 & 0.119 & 1.848 & 2.078 & 0.338 & 0.004 & 0.331 & 0.345 \\ 
        2035577729682322176 & 0.708 & 0.048 & 0.664 & 0.759 & 0.641 & 0.044 & 0.602 & 0.687 & 0.326 & 0.003 & 0.321 & 0.332 \\ 
        2067948245320365184 & 0.839 & 0.002 & 0.837 & 0.842 & 0.660 & 0.002 & 0.658 & 0.662 & 0.153 & 0.001 & 0.151 & 0.154 \\ 
        2129771310248902016 & 0.748 & 0.038 & 0.712 & 0.785 & 0.672 & 0.032 & 0.640 & 0.702 & 0.316 & 0.002 & 0.311 & 0.321 \\ 
        2185171578009765632 & 1.190 & 0.030 & 1.160 & 1.218 & 1.093 & 0.032 & 1.059 & 1.124 & 0.412 & 0.005 & 0.405 & 0.419 \\ 
        2198442167969655296 & 1.021 & 0.046 & 0.976 & 1.067 & 0.772 & 0.027 & 0.745 & 0.800 & 0.203 & 0.001 & 0.198 & 0.208 \\ 
        3283823387685219328 & 1.017 & 0.019 & 0.995 & 1.032 & 0.755 & 0.014 & 0.739 & 0.766 & 0.124 & 0.004 & 0.093 & 0.151 \\ 
        3312631623125272448 & 0.990 & 0.077 & 0.925 & 1.076 & 0.700 & 0.054 & 0.655 & 0.761 & 0.139 & 0.002 & 0.128 & 0.150 \\ 
        3366718833479009408 & 1.050 & 0.011 & 1.038 & 1.061 & 0.673 & 0.006 & 0.667 & 0.679 & 0.017 & 0.001 & 0.012 & 0.022 \\ 
        3409686270424363008 & 0.794 & 0.016 & 0.778 & 0.811 & 0.561 & 0.010 & 0.551 & 0.571 & 0.105 & 0.001 & 0.102 & 0.109 \\ 
        3427930123268526720 & 0.933 & 0.087 & 0.849 & 1.018 & 0.308 & 0.029 & 0.282 & 0.336 & 0.074 & 0.001 & 0.067 & 0.079 \\ 
        3536759371865789568 & 1.588 & 0.075 & 1.514 & 1.662 & 1.380 & 0.051 & 1.331 & 1.428 & 0.208 & 0.002 & 0.201 & 0.214 \\ 
        3549833939509628672 & 0.957 & 0.042 & 0.916 & 1.001 & 0.900 & 0.032 & 0.868 & 0.931 & 0.359 & 0.004 & 0.352 & 0.366 \\ 
        3931519127529822208 & 1.085 & 0.076 & 1.015 & 1.164 & 0.838 & 0.034 & 0.806 & 0.872 & 0.144 & 0.001 & 0.139 & 0.148 \\ 
        3935131126305835648 & 1.418 & 0.086 & 1.338 & 1.507 & 0.998 & 0.060 & 0.942 & 1.061 & 0.135 & 0.001 & 0.129 & 0.141 \\ 
        3954536956780305792 & 0.694 & 3.100 & 0.262 & 0.933 & 0.629 & 2.756 & 0.238 & 0.853 & 0.333 & 0.019 & 0.271 & 0.352 \\ 
        3964895043508685312 & 0.674 & 0.183 & 0.531 & 0.835 & 0.599 & 0.163 & 0.473 & 0.743 & 0.218 & 0.006 & 0.198 & 0.236 \\ 
        4145362250759997952 & 1.492 & 0.400 & 1.145 & 1.929 & 0.956 & 0.191 & 0.782 & 1.154 & 0.105 & 0.003 & 0.079 & 0.133 \\ 
        4228891667990334976 & 1.294 & 0.095 & 1.192 & 1.376 & 1.116 & 0.081 & 1.029 & 1.187 & 0.304 & 0.002 & 0.300 & 0.308 \\ 
        4354357901908595456 & 0.756 & 0.067 & 0.693 & 0.825 & 0.428 & 0.023 & 0.408 & 0.452 & 0.060 & 0.001 & 0.056 & 0.064 \\ 
        4589258562501677312 & 0.681 & 0.066 & 0.617 & 0.750 & 0.559 & 0.041 & 0.520 & 0.600 & 0.292 & 0.003 & 0.284 & 0.300 \\ 
        5762455439477309440 & 0.802 & 0.069 & 0.739 & 0.874 & 0.646 & 0.034 & 0.613 & 0.680 & 0.177 & 0.002 & 0.169 & 0.185 \\ 
        6244076338858859776 & 0.776 & 0.015 & 0.760 & 0.790 & 0.684 & 0.011 & 0.673 & 0.694 & 0.294 & 0.001 & 0.291 & 0.296 \\ 
        6799537965261994752 & 0.892 & 0.006 & 0.890 & 0.900 & 0.802 & 0.005 & 0.801 & 0.809 & 0.326 & 0.002 & 0.322 & 0.331 \\ \hline
    \end{tabular}
\end{table*}


\section{Jacobian calculations} \label{annex-error-jacobian-calculation}

The TMB algorithm provides in output the orbital parameters with their associated uncertainties, in particular the period $P$ with $\sigma_P$ and the parallax $\varpi$ with $\sigma_\varpi$. It additionally gives the semi-major axis of the primary orbit $a_1$ with its error $\sigma_{a1}$, the mass ratio $q$ with $\sigma_{q}$ and the flux fraction $\beta$ with $\sigma_\beta$. The global covariance matrix is also calculated.

Using $ a_2 = \frac{a_1}{q}$, the masses are derived with:

\begin{equation} \label{eq-annex-mass1-calcul}
    \mathcal{M}_1 = \frac{(a_1+a_2)^3}{P^2 ~(1+q)} = \frac{a_1^3 ~(1+ q)^2}{P^2 ~q^3}
\end{equation}
\begin{equation} \label{eq-annex-mass2-calcul}
    \mathcal{M}_2 = \frac{(a_1+a_2)^3 q}{P^2 ~(1+q)} = \frac{a_1^3 ~(1+ q)^2}{P^2 ~q^2}
\end{equation}

and the magnitudes are obtained through Eq.~\ref{absolute-magnitude-G-primary} and Eq.~\ref{absolute-magnitude-G-secondary} from the parameters ($\varpi, \beta$) as well as the apparent magnitude $G$ and the absorption $A_G$. The latest two are independent and independent of the other parameters. We therefore for simplicity consider here only the extinction corrected magnitude $G_0=G-A_G$ which has a variance $\sigma_{G_0}^2 = \sigma_G^2 + \sigma_{A_G}^2$.

The covariance of the parameters ($a_1, q, P, \varpi, \beta$) $\Sigma_{TMB}$ are provided by TMB. The full covariance of our input parameters ($a_1, q, P, \varpi, \beta, G_0$) is then
\begin{equation}
    \Sigma_{params} =  \left(
    \begin{array}{ccc}
        \Sigma_{TMB} & 0 \\
        0 & \sigma_{G_0}^2 \\
    \end{array}
    \right)
\end{equation}    

The covariance of our output parameters ($\mathcal{M}_1, \mathcal{M}_2, M_{G_1}, M_{G_2}$), $\Sigma$, is then 
$$\Sigma = J^T~\Sigma_{params}~J$$
with $J$ the Jacobian of the transformation from ($a_1, q, P, \varpi, \beta, G_0$) to ($\mathcal{M}_1, \mathcal{M}_2, M_{G_1}, M_{G_2}$):

\begin{equation} \label{eq-annex-jacobian}
    J =
    \left(
    \begin{array}{cccc}
        \partial_{a_1}\mathcal{M}_1 & \partial_{a_1}\mathcal{M}_2 & 0 & 0 \\
        \partial_{q}\mathcal{M}_1 & \partial_{q}\mathcal{M}_2 & 0 & 0 \\
        \partial_{P}\mathcal{M}_1 & \partial_{P}\mathcal{M}_2 & 0 & 0 \\
        0 & 0 & \partial_{\varpi}M_{G_1} & \partial_{\varpi}M_{G_2} \\
        0 & 0 & \partial_{\beta}M_{G_1} & \partial_{\beta}M_{G_2} \\
        0 & 0 & \partial_{G_0}M_{G_1} & \partial_{G_0}M_{G_2} \\
    \end{array}
    \right)
\end{equation}

The derivatives of the primary mass are given in the equations below :
\begin{equation} \label{eq-annex-dM1-da1}
    \partial_{a_1}\mathcal{M}_1 = \frac{3 a_1^2 ~(1+ q)^2}{P^2 ~q^3}
\end{equation}
\begin{equation} \label{eq-annex-dM1-dq}
     \partial_{q}\mathcal{M}_1 = \frac{a_1^3}{P^2} \left( \frac{2(1+ q)}{q^3} - \frac{3 (1+ q)^2}{q^4} \right)
\end{equation}
\begin{equation} \label{eq-annex-dM1-dP}
    \partial_{P}\mathcal{M}_1 = - 2 \frac{a_1^3 ~(1+ q)^2}{P^3 ~q^3}
\end{equation}

The derivatives of the secondary mass are given in the equations below :

\begin{equation} \label{eq-annex-dM2-da1}
    \partial_{a_1}\mathcal{M}_2 = \frac{3 a_1^2 ~(1+ q)^2}{P^2 ~q^2}
\end{equation}
\begin{equation} \label{eq-annex-dM2-dq}
    \partial_{q}\mathcal{M}_2 = \frac{a_1^3}{P^2} \left( \frac{2(1+ q)}{q^2} - \frac{2 (1+ q)^2}{q^3} \right)
\end{equation}
\begin{equation} \label{eq-annex-dM2-dP}
    \partial_{P}\mathcal{M}_2 = - 2 \frac{a_1^3 ~(1+ q)^2}{P^3 ~q^2}
\end{equation}

As a reminder, the absolute magnitudes are given by the Eq. \ref{absolute-magnitude-G-primary} and \ref{absolute-magnitude-G-secondary}.

The derivatives of the primary magnitude are given below :

\begin{equation} \label{eq-annex-dMG1-dplx}
    \partial_\varpi M_{G_1} = \frac{5}{\log(10)} \frac{1}{\varpi}
\end{equation}
\begin{equation} \label{eq-annex-dMG1-dbeta}
    \partial_\beta M_{G_1} = \frac{2.5}{\log(10)} \frac{1}{1 - \beta}
\end{equation}
\begin{equation} \label{eq-annex-dMG1-dG0}
    \partial_{G_0} M_{G_1} = 1
\end{equation}

The derivatives of the secondary magnitude are given below :
   
\begin{equation} \label{eq-annex-dMG2-dplx}
    \partial_\varpi M_{G_2} = \frac{5}{\log(10)} \frac{1}{\varpi}
\end{equation}
\begin{equation} \label{eq-annex-dMG2-dbeta}
    \partial_\beta M_{G_2} = - \frac{2.5}{\log(10)} \frac{1}{\beta}
\end{equation}
\begin{equation} \label{eq-annex-dMG2-dG0}
    \partial_{G_0} M_{G_2} = 1
\end{equation}

\end{document}